\documentstyle[epsfig]{aipproc}

\begin{document}
\title{Can we distinguish an MSSM Higgs from a SM Higgs at a Linear 
Collider?\thanks{Talk presented at the Linear Collider Workshop 2000
(LCWS2000), 
24-28 October 2000, Fermilab, Batavia, Illinois, USA.}}

\author{Heather E. Logan\thanks{Electronic address: logan@fnal.gov}}
\address{Theoretical Physics Department \\ 
Fermi National Accelerator Laboratory, Batavia, IL 60510-0500, 
USA\thanks{Fermilab is operated by Universities Research Association Inc.\
under contract no.~DE-AC02-76CH03000 with the U.S. Department of
Energy.}}

\maketitle

\begin{abstract}
We study the prospects for distinguishing the CP-even Higgs boson of the 
minimal supersymmetric extension of the Standard Model (MSSM)
from the Standard Model (SM) Higgs boson 
by measuring its branching ratios at an $e^+e^-$
linear collider.  
The regions of the $M_A - \tan\beta$ plane
in which an MSSM Higgs boson can be distinguished from 
the SM Higgs boson depend strongly upon the supersymmetric
parameters that enter 
the radiative corrections to the Higgs mass matrix and 
the Higgs couplings to fermions.  
In some regions of parameter space it is possible
to extract the supersymmetric correction to the relation between the $b$ 
quark mass and its Yukawa coupling from Higgs branching ratio measurements.
\end{abstract}

\vskip-12.5cm
\noindent
FERMILAB-Conf-00/324-T \hfill hep-ph/0012202
\vskip12cm
\noindent
Present knowledge of the radiative corrections to
the Higgs sector of the 
minimal supersymmetric extension of the Standard Model (MSSM)
allows one to compute the branching
ratios (BRs) of the MSSM Higgs bosons with high precision.  
The BRs of the Standard Model (SM)-like Higgs boson of the MSSM 
(i.e., the MSSM Higgs boson with the largest couplings to $WW$ and $ZZ$,
denoted $H_{MSSM}$)
in general differ from those of a SM Higgs boson (denoted $H_{SM}$) 
of the same mass.  
If these BRs are measured to high enough precision, 
they can be used to distinguish between $H_{SM}$ and $H_{MSSM}$.
In this talk we examine the potential of Higgs BR measurements
at a future $e^+e^-$ linear collider (LC) to distinguish 
$H_{MSSM}$ from $H_{SM}$ in various regions of MSSM
parameter space that give rise to significantly different behaviors
of the MSSM Higgs bosons.  For the details of our analysis 
see Ref.~\cite{CHLM}.

At tree level, the MSSM Higgs sector depends on only two
parameters, $M_A$ and $\tan\beta$.
Radiative corrections to the Higgs mass matrix and vertex 
corrections to the Higgs-fermion Yukawa couplings 
introduce significant dependence on other MSSM parameters
(for a review and references see Ref.~\cite{HiggsWGrep}).
The radiative corrections to the Higgs mass matrix lead to corrections
to the mixing angle $\alpha$ for the two CP-even MSSM Higgs bosons, which
affect the Higgs couplings to fermions and vector bosons.  The vertex
corrections to the Higgs-fermion Yukawa couplings (denoted $\Delta_b$
for $b$ quarks) primarily modify the Higgs couplings to $b \bar b$
and depend on the parameters $\mu M_{\tilde g}$ and $\mu A_t$ and the 
squark masses.
Explicit formulae may be found in Refs.~\cite{CHLM,HiggsWGrep}.

\begin{figure}[t]
	\resizebox{\textwidth}{!}
	{\includegraphics{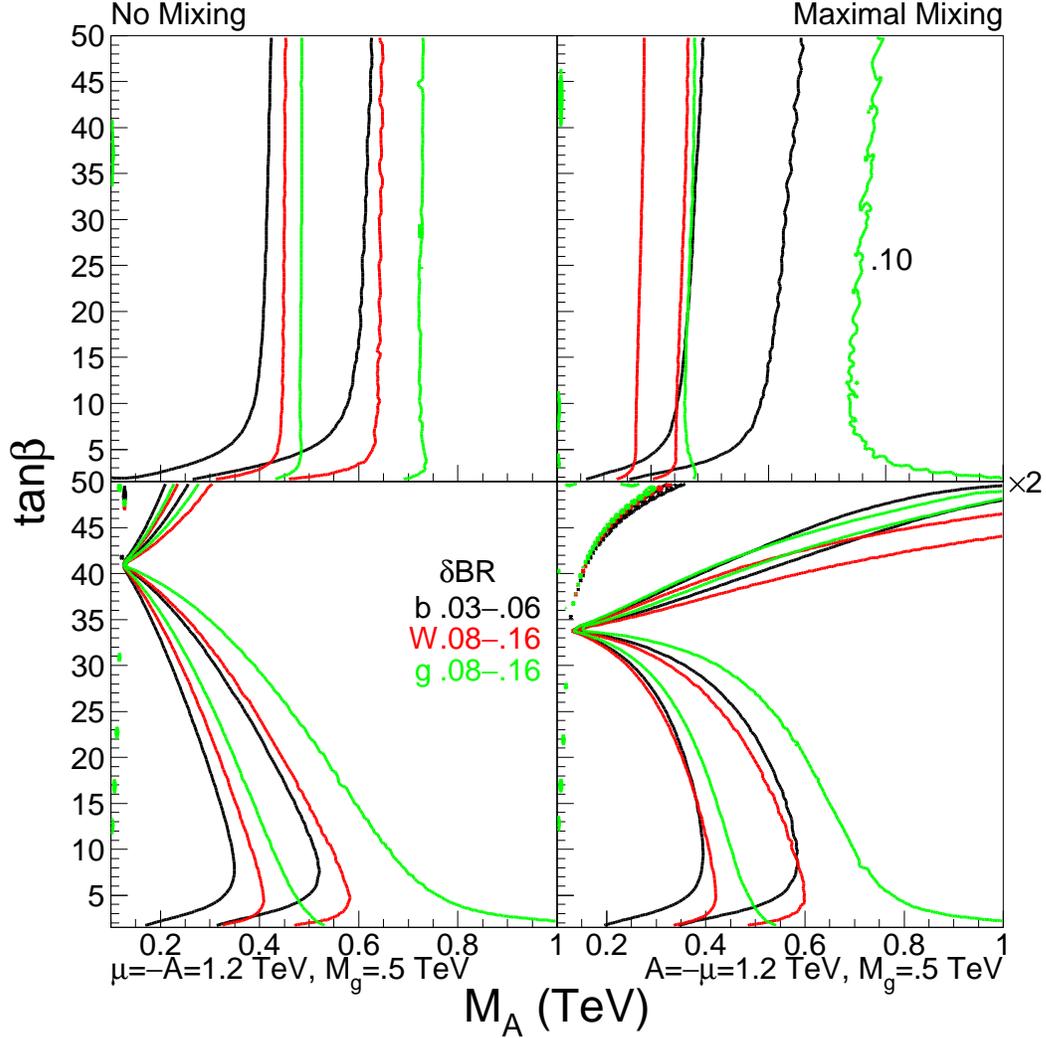}}
	\caption{The $1 \sigma$ and $2 \sigma$ contours of 
$\delta BR(b)$ (black), $\delta BR(W)$ 
(red or dark gray) and $\delta BR(g)$ (green or light gray) 
in the no mixing scenario (top left),
the maximal mixing scenario (top right),
and the large $\mu$ and $A_t$ scenario with  
$\mu = -A_t = 1.2$ TeV (bottom left) and 
$\mu = -A_t = -1.2$ TeV (bottom right).  
In the maximal mixing scenario (top right) we plot $M_A$ between 0.1 and
2 TeV and $\delta BR(g) = 0.16$ and 0.10 (here $\delta BR(g) = 0.08$
lies above $M_A = 2$ TeV).}
	\label{fig:nomixing}
\end{figure}

We examine three benchmark scenarios
(Table~\ref{tab:scenarios}) that lead to very different behaviors 
for $H_{MSSM}$.
\begin{table} 
	\caption{MSSM parameters in TeV for the three benchmark 
scenarios.  We set the gaugino mass parameter $M_2 = 0.2$ TeV.}
	\label{tab:scenarios}
	\begin{tabular}{c c c c c c} 
	Benchmark & $\mu$ & $X_t \equiv A_t - \mu \cot\beta$ & $A_b$
		& $M_S$ & $M_{\tilde g}$ \\
	\hline
	No Mixing & $-0.2$ & 0 & $A_t$ & 1.5 & 1.0 \\
	Maximal Mixing & $-0.2$ & $\sqrt{6} M_S$ & $A_t$ & 1.0 & 1.0 \\
	Large $\mu$ and $A_t$ & $\pm 1.2$ & $\mp 1.2(1 + \cot\beta)$
		& 0 & 1.0 & 0.5 \\
	\hline
	\end{tabular}
\end{table}
These scenarios are chosen so that the Higgs mass is above its present
upper bound from LEP and to maximize the effect of the choice of MSSM 
parameters on the behavior of the Higgs BRs.
We use the program HDECAY \cite{HDECAY} to which we have added 
the Yukawa vertex corrections.
In each of the benchmark scenarios
we compute the mass and BRs of $H_{MSSM}$
at each point in the $M_A - \tan\beta$ plane.  We compare
the BRs of $H_{MSSM}$ to those of $H_{SM}$ with the 
same mass and plot contours of
$\delta BR \equiv |BR_{MSSM} - BR_{SM}|/BR_{SM}$.  
In Table~\ref{tab:BRmeas} we show the expected uncertainties of 
BR measurements 
at a LC for a 120 GeV SM Higgs boson
from Refs.~\cite{VanKooten} ($\sqrt{s} = 500$ GeV with 200 fb$^{-1}$)
and \cite{Battaglia} ($\sqrt{s} = 350$ or 500 GeV with 500 fb$^{-1}$).
\begin{table}
	\caption{Expected fractional uncertainty of BR measurements at a LC
for a 120 GeV SM Higgs boson.}
	\label{tab:BRmeas} 
	\begin{tabular}{r c c c c c c}
	 & $b \bar b$ & $WW^*$ & $\tau^+ \tau^-$ 
		& $c \bar c$ & $gg$ & $\gamma\gamma$ \\ \hline
	Ref.~\cite{VanKooten} & 0.03 & 0.08 & 0.07 & 0.15 & 0.08 & 0.22 \\
	Ref.~\cite{Battaglia} & 0.024 & 0.054 & 0.083--0.135 
		& -- & 0.055 & -- \\
	\hline 
	\end{tabular}
\end{table}
In Fig.~\ref{fig:nomixing} we plot the $1 \sigma$ and $2 \sigma$ contours
(based on the uncertainties from Ref.~\cite{VanKooten} 
(Table~\ref{tab:BRmeas}))
of $\delta BR(b)$, $\delta BR(W)$ and $\delta BR(g)$
in the three benchmark scenarios.

\renewcommand{\topfraction}{.99}
In the top left panel of Fig.~\ref{fig:nomixing} 
we examine
the no mixing scenario.
In this scenario the reach in $M_A$ for distinguishing $H_{MSSM}$ 
from $H_{SM}$ is fairly independent of $\tan\beta$.
With the uncertainties 
in Ref.~\cite{VanKooten} (Table~\ref{tab:BRmeas}), $BR(g)$ gives the
greatest reach in $M_A$, allowing one to distinguish $H_{MSSM}$
from $H_{SM}$ at $1 \sigma$ ($2 \sigma$) for $M_A \lesssim 725$ GeV
(475 GeV).

In the top right panel of Fig.~\ref{fig:nomixing} 
we examine
the maximal mixing scenario.
In this scenario we find significant deviations in $BR(b)$ and $BR(g)$ 
from their SM values even at very large $M_A > 1$ TeV.  
We find that one can distinguish $H_{MSSM}$ from $H_{SM}$ at $1 \sigma$
using $\delta BR(g)$ even for $M_A \simeq 2$ TeV, while at $2 \sigma$
the reach in $\delta BR(g)$ and $\delta BR(b)$ are comparable
and one can distinguish $H_{MSSM}$ from $H_{SM}$ for 
$M_A \lesssim 650$ GeV.

In the two bottom panels of Fig.~\ref{fig:nomixing} 
we examine
the large $\mu$ and $A_t$ scenario.
In this scenario we find that at large $\tan\beta$ there are 
regions of parameter space in which $H_{MSSM}$ cannot be 
distinguished from $H_{SM}$ even for very low values of 
$M_A \simeq 200$ GeV.  
Thus the regions of the $M_A - \tan\beta$ plane
in which $H_{MSSM}$ can be distinguished from 
$H_{SM}$ depend strongly on the supersymmetric parameters.

Finally, in Ref.~\cite{CHLM} we show that it is possible to extract $\Delta_b$ 
from measurements of ratios of 
branching ratios: $\Delta_b = (1 - \sqrt{x})/(\sqrt{x} - \sqrt{y})$
with $x = (BR(b)/BR(\tau))/(BR(b)/BR(\tau))_{SM}$ and 
$y = (BR(c)/BR(\tau))/(BR(c)/BR(\tau))_{SM}$.
\begin{figure}
	\resizebox{\textwidth}{!}
	{\includegraphics{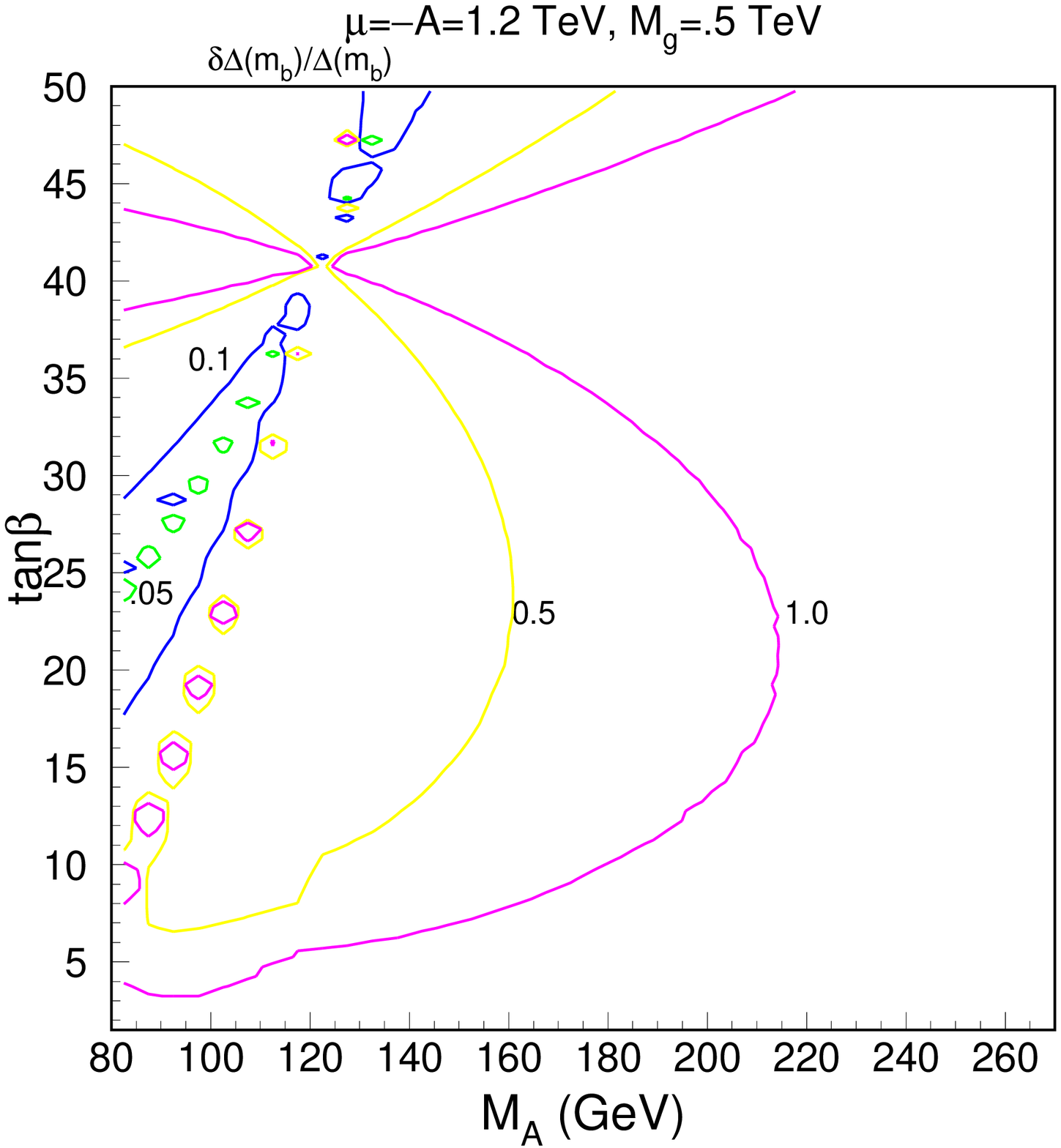}\includegraphics{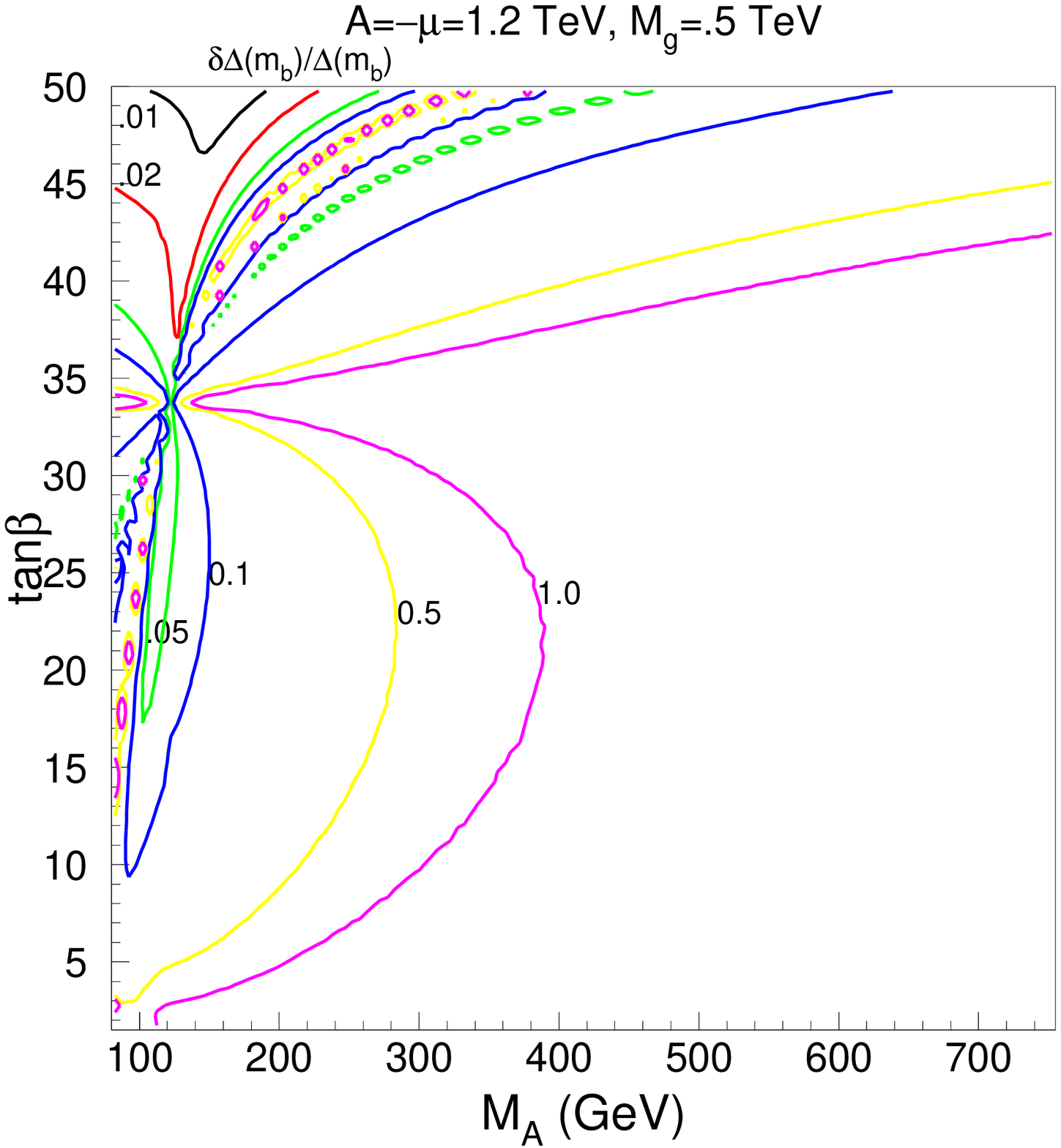}}
	\caption{Contours of the fractional error in the 
determination of $\Delta_b$ in the large $\mu$ and $A_t$ scenario.  Here 
$\mu = -A_t = 1.2$ TeV (left) 
and $\mu = -A_t = -1.2$ TeV (right).}
	\label{fig:dmb}
\end{figure}
In Fig.~\ref{fig:dmb} we show the fractional error in the determination
of $\Delta_b$ from measurements of $BR(b)/BR(\tau)$ and $BR(c)/BR(\tau)$
in the large
$\mu$ and $A_t$ scenario (see Table~\ref{tab:scenarios}),
in which $\Delta_b$ is quite sizeable.
We assume BR uncertainties as in Ref.~\cite{VanKooten}.
Note that for $\mu > 0$ (the left panel of Fig.~\ref{fig:dmb}),
$\Delta_b$ can only be distinguished from zero for
moderate to large $\tan\beta$ and $M_A \lesssim 170$ GeV.
In contrast, for $\mu < 0$
(the right panel of Fig.~\ref{fig:dmb}),
$\Delta_b$ can be determined with 10\% accuracy even for $M_A$ as 
large as 600 GeV for large $\tan\beta$.
This measurement of $\Delta_b$ may ultimately be combined with other 
measurements to determine the underlying SUSY parameters.

I thank M.~Carena, H.~E.~Haber and S.~Mrenna, with whom this work was 
performed.

\vspace{-0.5cm}

\end{document}